\documentclass[reprint,prl,aps,superscriptaddress]{revtex4-2}

\usepackage{times}
\usepackage{graphicx}
\usepackage{amsmath, braket, amsfonts}
\usepackage{amssymb}
\usepackage{bm, color, ulem}
\usepackage{xcolor}
\usepackage{ragged2e}
\usepackage{wrapfig,lipsum,booktabs}
\usepackage{lineno}

\newcommand{\UCSB}{Department of Physics, University of California at Santa Barbara, Santa Barbara, California 93106, USA}
\newcommand{\nimsTT}{Research Center for Materials Nanoarchitectonics, National Institute for Materials Science, 1-1 Namiki, Tsukuba 305-0044, Japan}
\newcommand{\nimsKW}{
Research Center for Electronic and Optical Materials, National Institute for Materials Science, 1-1 Namiki, Tsukuba 305-0044, Japan}

\usepackage{siunitx}

\newcommand{\be}{\begin{equation}}
\newcommand{\ee}{\end{equation}}

\usepackage{mathtools}
\usepackage{physics}

\DeclareUnicodeCharacter{03BD}{{$\nu$}}
\AtBeginDocument{\RenewCommandCopy\qty\SI}
\AtBeginDocument{\RenewCommandCopy\qty\SI} 
\begin{document}

\title{Superconductivity and quantized anomalous Hall in rhombohedral graphene}

\author{Youngjoon Choi}
\email{These authors contributed equally} 
\author{Ysun Choi}
\email{These authors contributed equally} 
\author{Marco Valentini}
\email{These authors contributed equally} 
\author{Caitlin L. Patterson}
\author{Ludwig F. W. Holleis}
\author{Owen I. Sheekey}
\author{Hari Stoyanov}
\author{Xiang Cheng} 
\affiliation{\UCSB}
\author{Takashi Taniguchi}
\affiliation{\nimsTT}
\author{Kenji Watanabe}
\affiliation{\nimsKW}
\author{Andrea F. Young}
\email{andrea@physics.ucsb.edu}
\affiliation{\UCSB}
\date{\today}

\maketitle 

\textbf{
Inducing superconducting correlations in chiral edge states is predicted to generate topologically protected zero energy modes with exotic quantum statistics~\cite{qi_chiral_2010,clarke_exotic_2013,clarke_exotic_2014,mong_universal_2014,lian_topological_2018,lindner_fractionalizing_2012}.  
Experimental efforts to date have focused on engineering interfaces between superconducting materials--typically amorphous metals---and semiconducting quantum Hall ~\cite{lee_inducing_2017,amet_supercurrent_2016,vignaud_evidence_2023,gul_andreev_2022,barrier_one-dimensional_2024} 
or quantum anomalous Hall (QAH) ~\cite{uday_induced_2024,
atanov_proximity-induced_2024} systems. 
However, the strong interfacial disorder inherent in this approach can prevent the formation of isolated topological modes~\cite{tang_vortex-enabled_2022,kurilovich_disorder-enabled_2023,
kurilovich_criticality_2023,ji_eh_2018}. 
An appealing alternative is to use low-density flat band materials where the ground state can be tuned between intrinsic superconducting and quantum anomalous Hall states using only the electric field effect.  
However, quantized transport and superconductivity have not been simultaneously achieved. 
Here, we show that rhombohedral tetralayer graphene aligned to a hexagonal boron nitride substrate hosts a quantized anomalous Hall state at superlattice filling $\bm{\nu=-1}$ as well as a superconducting state at $\bm{\nu\approx -3.5}$ at zero magnetic field.  
Remarkably, gate voltage can also be used to actuate nonvolatile switching of the chirality in the quantum anomalous Hall state~\cite{polshyn_electrical_2020}, allowing, in principle, arbitrarily reconfigurable networks of topological edge modes in locally gated devices.  Thermodynamic compressibility measurements further reveal a topologically ordered fractional Chern insulator at $\bm{\nu=2/3}$~\cite{lu_fractional_2024}---also stable at zero magnetic field---enabling proximity coupling between superconductivity and fractionally charged edge modes.  Finally, we show that, as in rhombohedral bi- and trilayers~\cite{zhang_enhanced_2023,patterson_superconductivity_2024,yang_diverse_2024}, integrating a transition metal dichalcogenide layer to the heterostructure nucleates a new superconducting pocket~\cite{zhang_enhanced_2023,holleis_nematicity_2024,li_tunable_2024,patterson_superconductivity_2024,yang_diverse_2024}, while leaving the topology of the $\bm{\nu=-1}$ quantum anomalous Hall state intact.  
Our results pave the way for a new generation of hybrid interfaces between superconductors and topological edge states in the low-disorder limit.}

\begin{figure*}
    \centering
    \includegraphics[width=183 mm]{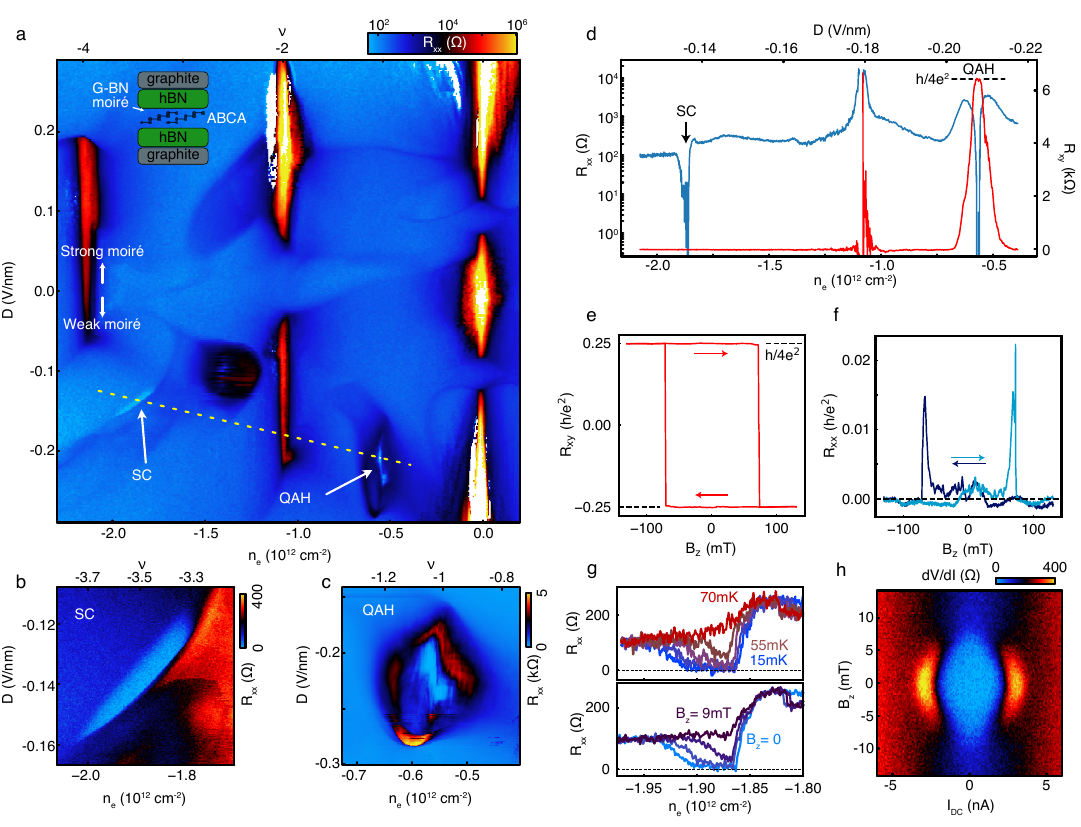}
    \caption{\textbf{Gate-tunable superconductivity and quantized anomalous Hall effect in hBN aligned rhombohedral tetralayer graphene.} 
    \textbf{a}, $R_{xx}$ as a function of $n_e$ and $D$.  Zero resistance superconductivity (SC) and quantized anomalous Hall (QAH) states are indicated. Inset: device schematic, showing rhombohedral tetralayer with ABCA stacking order, encapsulating hexagonal boron nitride (hBN) layers and graphite gates. The crystal axes of the top hBN and graphene tetralayer are aligned, so that hole carriers occupy the layer with strong moir\'e potential for $D>0$, as indicated (See Methods and Fig.~\ref{fig:optical} for the device details).
    \textbf{b}, \textbf{c}, Detail of $R_{xx}$ around the the superconducting pocket (\textbf{b}) and the QAH region at $\nu=-1$ (\textbf{c}). \textbf{d}, $R_{xx}$ and $R_{xy}$, measured along the trajectory indicated in panel \textbf{a}.  
    \textbf{e}, \textbf{f}, $R_{xy}$ and $R_{xx}$ within the $\nu=-1$ quantum anomalous Hall plateau measured as a function of $B_z$ at $n_e = -0.564 \times10^{12} \,\si{cm^{-2}}$, $D = -0.208 \,\si{V/nm}$.
    \textbf{g}, $T$ and $B_z$ dependence of $R_{xx}$ in the superconducting pocket at $D$ = $-0.138 \,\si{V/nm}$. 
    \textbf{h}, Nonlinear resistivity measured as a function of $B_z$ and $I_{DC}$ in the superconducting pocket at $n_e = -1.879 \times10^{12} \,\si{cm^{-2}}$, $D = -0.138 \,\si{V/nm}$.}
    \label{fig:1}
\end{figure*}

\section{Introduction}
The discovery of superconductivity in 1911 and the quantized Hall effect~\cite{klitzing_new_1980} in 1980 together have come to define the two dominant paradigms--based on symmetry breaking and on topology, respectively---for understanding condensed matter systems. 
In recent years, there has been intense interest in the possibility of introducing superconducting pairing to the chiral edge modes of the quantum Hall state, as this combination is predicted to give rise to new modes with unconventional quantum statistics~\cite{qi_chiral_2010,clarke_exotic_2013,mong_universal_2014,lian_topological_2018,lindner_fractionalizing_2012}, and allow for the creation of electronic devices with novel functionality~\cite{clarke_exotic_2014}. 
Practically, however, interfacing superconductivity with chiral edge modes is rendered challenging by the incompatibility of materials hosting these paradigmatic zero-resistance states: superconductivity is most common in high density metals, while quantized Hall effects typically occur in low carrier density semiconducting materials. As a result, interpretation of experimental data has had to contend with the effects of disorder arising from the interface between different materials. 
Two dimensional flat band systems such as twisted bilayer graphene, 
transition metal dichalcogenide homo- and heterobilayers, and rhombohedral graphene multilayers provide the opportunity to circumvent this challenge. In all of these systems, a high density of states provides the setting for intrinsic superconductivity and magnetism, while the Berry curvature native to honeycomb materials generically leads to topological bands. 
However, to date no single device has shown both quantized anomalous Hall transport and intrinsic superconductivity at zero magnetic field. 
In twisted bilayer graphene, gate tunability between a superconductor and magnetic state was demonstrated, but without quantized edge state transport~\cite{stepanov_competing_2021}. 
In twisted WSe$_2$, thermodynamic signatures of quantum anomalous Hall states were observed~\cite{foutty_mapping_2024}, but in a different regime of twist angle than superconductivity~\cite{xia_unconventional_2024,guo_superconductivity_2024}. Other TMD systems such as twisted MoTe$_2$ and MoTe$_2$/WSe$_2$ show integer and fractionally quantized edge transport~\cite{li_quantum_2021,park_observation_2023}, but no superconductivity has been reported. 

Rhombohedral graphene multilayers provide an appealing alternative due to their low disorder and high experimental reproducibility.
Absent crystallographic alignment to an hBN cladding layer, superconductivity and magnetic metallic states have been observed in both bi- and trilayer systems~\cite{zhou_half-_2021,zhou_superconductivity_2021,zhou_isospin_2022,zhang_enhanced_2023,li_tunable_2024,patterson_superconductivity_2024,yang_diverse_2024}.  
Integrating a moir\'e potential via hBN alignment allows for quantized anomalous Hall states at finite carrier density via commensuration with the moir\'e superlattice potential. 
Signatures of a QAH state were first observed in aligned trilayers~\cite{chen_tunable_2020}, albeit without quantization at zero magnetic field. 
Signatures of superconductivity were also reported in this system~\cite{chen_signatures_2019}; however, the resistance of this state saturated to several hundred ohms at low temperature, and was resilient to perpendicular magnetic fields as large as 1T---features which subsequent work has shown to be uncharacteristic of rhombohedral graphene superconductors~\cite{zhou_superconductivity_2021,zhou_isospin_2022,zhang_enhanced_2023,holleis_nematicity_2024,patterson_superconductivity_2024,yang_diverse_2024}.  More recently, both integer and fractional QAH states were reported in aligned pentalayers~\cite{lu_fractional_2024} and hexalayers~\cite{xie_even-_2024}.  These results motivate the search for superconductivity in aligned rhombohedral multilayers.

\section{Superconductivity and QAH effect}

Here, we report the observation of superconductivity and quantized anomalous Hall effect in hBN aligned rhombohedral tetralayer graphene. 
Fig.~\ref{fig:1}a shows the carrier density-($n_e$) and displacement field ($D$) dependent longitudinal resistance ($R_{xx}$) of Device A, measured for  magnetic field $B=0$ and temperature $T = 15 \,\si{mK}$.    
High resistance features are observed at $|n_e| = (2.15 \pm 0.05) \times10^{12} \,\si{cm^{-2}}$, which we associate with full filling  ($\nu=\pm4$) of the moir\'e unit cell~\cite{chen_evidence_2019} giving an estimated moir\'e lattice constant of $\lambda = 14.5-14.8 \,\si{nm}$.  
Here we focus on two distinct pockets where $R_{xx}=0$ observed for $D<0$ where the carriers are polarized away from the aligned hBN interface (see Fig.~\ref{fig:1}b-c).
Fig.~\ref{fig:1}d shows $R_{xx}$ and $R_{xy}$ measured along the trajectory indicated in Fig.~\ref{fig:1}a. $R_{xx}$ drops below $10\,\si{\Omega}$ within each pocket; the two states show contrasting $R_{xy}$, which vanishes for the high hole-density pocket near $n_e\approx-1.9 \times 10^{12} \,\si{cm^{-2}}$ (consistent with a superconducting state) but is quantized to $R_{xy}=h/4e^{2}$ at $n_e\approx-0.6 \times10^{12} \,\si{cm^{-2}}$, corresponding to $\nu=-1$ moir\'e filling.  
$B_z$ dependent measurements within this state indeed show hysteretic switching between quantized values of $\pm h/4e^{2}$ (Fig.~\ref{fig:1}e), concomitant with $R_{xx}<.01 h/e^2$ (Fig.~\ref{fig:1}f), consistent with a quantized anomalous Hall state with Chern number $|C|=4$\, as theoretically predicted \cite{zhang_spontaneous_2011,park_topological_2023}. 
Temperature dependent measurements at this filling show behavior consistent with a magnetic Curie temperature of $4.5 \,\si{K}$ and a thermal activation gap of $\Delta\approx 10 \,\si{K}$ (see Fig.~\ref{fig:CI_energetics}).  

To confirm the identification of the superconducting phase, we study the temperature, current, and perpendicular magnetic field dependence of the zero-resistance state in Figs.~\ref{fig:1}g-h. Superconductivity is suppressed for temperatures $T>T_C\approx 55 \,\si{mK}$ and perpendicular magnetic field  $B_z>B_{c} \approx 7 \,\si{mT}$, and shows a critical current of $I_C\approx 2 \,\si{nA}$. 
These observations are in-line with existing work on unaligned rhombohedral trilayers~\cite{zhou_superconductivity_2021} and Bernal bilayers~\cite{zhou_isospin_2022} encapsulated in hexagonal boron nitride.  
Further evidence for superconductivity in the high-density pocket is a flux-tuned oscillation in the critical current observed for certain dopings, confirming the macroscopic phase coherence of the superconducting state (see Fig.~\ref{fig:SC_properties}).
Notably, the domain of superconductivity is reminiscent of hole-doped rhombohedral trilayer graphene, where superconductivity arises in a narrow strip bounded at lower $|n_e|$ by a sharp phase boundary likely associated with isospin symmetry breaking~\cite{zhou_superconductivity_2021}. 
An additional superconducting state is also observed near $\nu = -2$ on the strong moir\'e side, with a critical temperature $T_c \approx 90 \ \si{mK}$, described in Extended Data Fig.~\ref{fig:spin_pol}).

\section{Controllable switching of QAH edge chirality}

\begin{figure*}
    \centering
    \includegraphics[width=183mm]{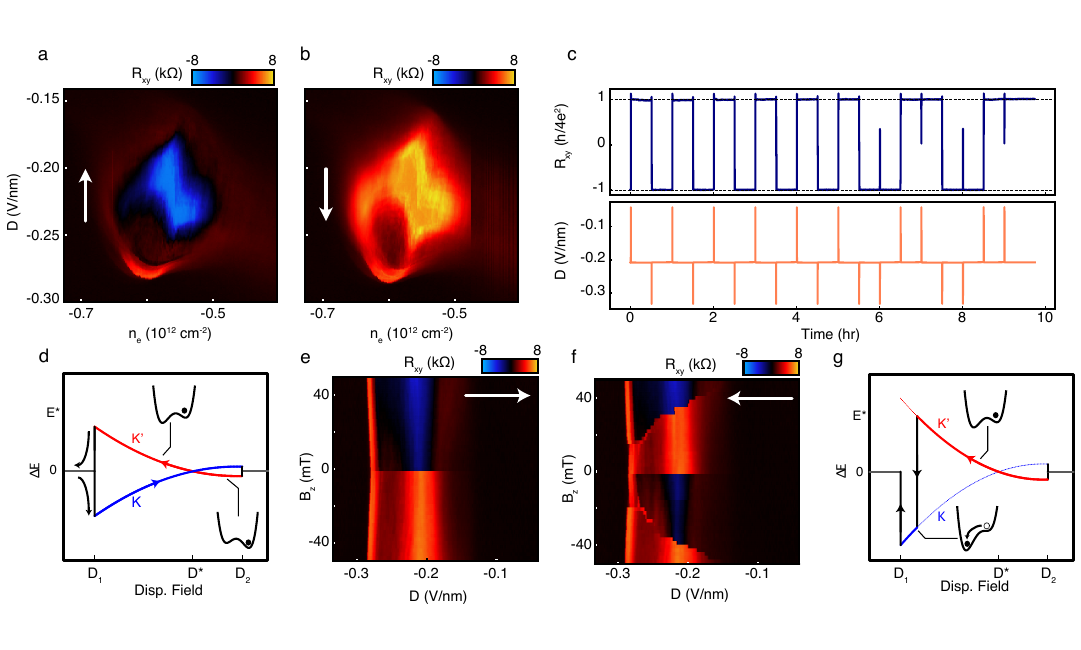}
    \caption{
    \textbf{Electric switching of edge mode chirality.}
    \textbf{a}, $R_{xy}$ measured as a function of $n_e$ and $D$ with $B_\parallel = 10 \,\si{mT}$ near $\nu=-1$. $D$ is swept from low to high, and $n_e$ incremented between sweeps.  
    \textbf{b}, The same as panel \textbf{a}, with $D$ swept high to low. 
    \textbf{c}, Time dependence of $R_{xy}$ (top panel) during a series of rapid excursion in $D$ (bottom panel). 
    \textbf{d}, Schematic diagram of $\Delta E=-B m_{K(K')}$ at low $B_z$. Along the $D$-tuned trajectory, the system goes from valley unpolarized to a valley polarized phase at $D_1$; within the valley polarized phase, $m_K=-m_{K'}$ changes sign at $D^*$; and the system undergoes a second transition to a valley unpolarized phase at $D=D_2$. 
    Upon entry to the polarized phase, the system polarizes into the low energy valley, but due to the finite energy barrier the system remains trapped in this valley even after it no longer corresponds to the ground state.  This results in opposite valley polarization throughout the valley polarized phase for opposite sweep directions.  Insets show schematics of the free energy  as a function of valley polarization $n_k$. 
    \textbf{e}, $R_{xy}$ as a function of $D$ and $B_z$ with fixed $B_\parallel=10 \,\si{mT}$.  $D$ is swept low to high. 
    \textbf{f},  The same as panel \textbf{e}, with $D$ swept high to low. 
    \textbf{g}, Energy level schematic at higher applied $B_z$.  The system relaxes at $\Delta E=E^*$, where the barrier vanishes.}
    \label{fig:2}
\end{figure*}

Figs.~\ref{fig:2}a-b show $R_{xy}$ measured in the vicinity of the $\nu=-1$ quantum anomalous Hall state at a nominal applied magnetic field $B_z\approx 0$ and $B_\parallel=10 \,\si{mT}$.  
For these measurements, $D$ is the fast axis, with $n_e$ stepped between traces from left to right. 
Figs.~\ref{fig:2}a and \ref{fig:2}b differ only in the direction of the $D$-axis sweep.  Both show quantized transport, but with opposite signs of quantized $R_{xy}$ at the same $n_e$ and $D$ values, implying an electric field induced reversal of the valley polarization. 
The switching effect is nonvolatile.  As shown in Fig.~\ref{fig:2}c, the chirality can be controllably reversed by excursions in the applied $D$, with the value of $R_{xy}$ remaining stable between excursions when $D$ is returned to the same value. 

Electric field control of magnetic switching has been reported previously in twisted and rhombohedral graphene multilayers~\cite{polshyn_electrical_2020,grover_chern_2022,han_orbital_2023,su_generalized_2024,sha_observation_2024}. 
Switching can also arise in metallic states due purely to the density-dependent orbital magnetization of Bloch states with finite Berry curvature, as occurs in the current sample near $\nu=-2.5$  (see Fig.~\ref{fig:AH_nu-2.5}). 
The key microscopic requirement is that the magnetization changes sign within a phase with fixed valley polarization~\cite{zhu_voltage-controlled_2020}. 
Time reversal symmetry mandates that the magnetic moment of states in opposite valleys have opposite signs ($m_{K'}=-m_K$), 
and at zero magnetic field, the valley polarization in an orbital magnet is chosen spontaneously. 
Indeed, entering a valley polarized phase at $B_z=0$ often results in switchy behavior associated with the randomly chosen valley polarization, visible in Figs.~\ref{fig:1}a, c, and \ref{fig:QAH_nD_stability}a.
At small but finite $B_z$, the valley degeneracy is lifted, introducing a valley splitting $\Delta E=2m_K B_z$ between the two valleys. Electrons polarized within the $K$ valley will only be in the ground state for $m_K>0$; if $m_K$ becomes negative, a $K$-polarized state will become metastable, and may persist in the presence of an energy barrier separating $K$ and $K'$ polarizations. 

Notably, in the present system switching is effective only when the $D$ range is sufficiently large to exit and reenter the valley polarized phase, with no relaxation observed for parameter sweeps \textit{within} the valley polarized phase at low $B_z$.  
This behavior can be explained if, upon entry to the valley polarized phase, the system polarizes in the low energy valley, but the low energy valley switches from K to K' within the valley polarized phase. 
Fig.~\ref{fig:2}d shows a schematic energy level diagram along with the trajectory of the system for rising and falling $D$. 
A finite energy barrier prevents relaxation throughout the valley polarized phase, resulting in the stable switching observed.  Interestingly, reliable nonvolatile switching does not occur for $B_\parallel=0$ (see Fig.~\ref{fig:inplane_switching}), suggesting that $B_\parallel$ suppresses partially valley polarized phases along the gate tuned trajectory in which relaxation barriers are small. 

As expected for magnetic systems, bistable behavior is eventually destroyed by sufficiently large $B_z$.  
Figs.~\ref{fig:2}e-f show $R_{xy}$ measured as a function of $D$ and $B_z$. Above a threshold value of $B_z$, the bistable regime shrinks and eventually vanishes at $B_z\approx 50 \,\si{mT}$. 
Fig.~\ref{fig:2}g describes a mechanism for this suppression.  
In the absence of thermal relaxation, the system remains in a metastable valley as long as the barrier is finite. Applied magnetic field destroys the barrier when $\Delta E$ becomes comparable to the barrier height $E^*$, allowing the system to relax to the ground state.

\section{Fractional Chern insulator at zero field}

\begin{figure*}
    \centering
    \includegraphics[width=183mm]{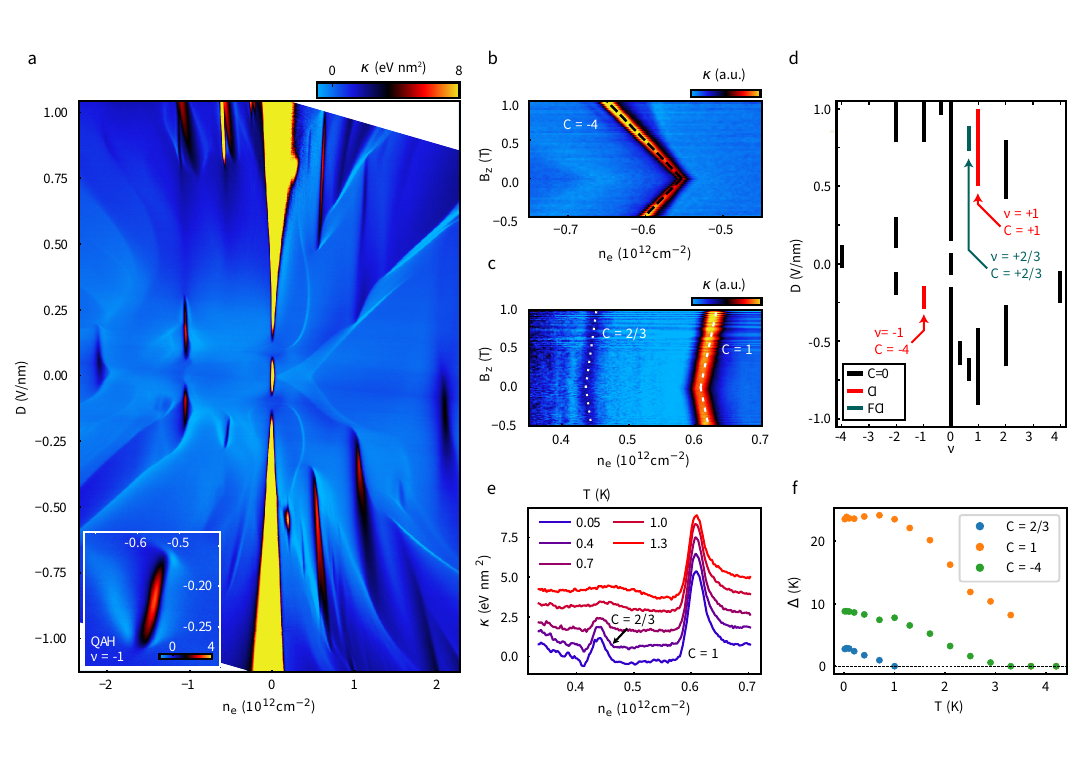}
    \caption{\textbf{Integer and fractional Chern insulators revealed by inverse compressibility measurements.} 
    \textbf{a}, $n_e$- and $D$-dependent inverse compressibility $\kappa = \partial \mu/ \partial n_e$. Gapped states manifest as peaks in $\kappa$ at fixed commensurate filling.  Note that constant $\nu$ contours are not vertical on this scale, due to the quantum capacitance of the strong insulator at the charge neutrality.  
    Inset: $\kappa$ near the $\nu=-1$ QAH state. 
    \textbf{b}, magnetic field dependence of $\kappa$ near the $\nu=-1$ QAH state measured at $D = -0.208 \,\si{V/nm}$. The Streda slope is consistent with $C=-4$. 
    \textbf{c}, Magnetic field dependence of $\kappa$ for $\nu=1$ and $2/3$ at $D = 0.885 \,\si{V/nm}$. Their slope difference is consistent with a $C=1$ and $C=2/3$, respectively. See Fig.~\ref{fig:slope_extraction}. 
    \textbf{d}, Classification of correlated incompressible peaks shown in \textbf{a}.  
    \textbf{e}, Temperature dependence of $\kappa$ for $C=2/3$ and $C=1$ Chern insulators at $D = 0.885 \,\si{V/nm}$. $C=2/3$ state is suppressed for $T\approx 1 \,\si{K}$. Curves are offset by $1 \,\si{eV\,nm}^2$ for clarity.
    \textbf{f}, Thermodynamic gap extracted from $\kappa$ (see Fig.~\ref{fig:CI_energetics}) as a function of temperature for $C=2/3, 1$ and $-4$ Chern insulators of panels \textbf{b} and \textbf{c}.}
    \label{fig:3}
\end{figure*}

Signatures of topologically nontrivial gapped phases are also evident in measurements of the inverse compressibility, $\kappa=\partial \mu/\partial n_e$, shown in Fig.~\ref{fig:3}a.  
Numerous peaks in the inverse compressibility corresponding to correlated insulator states are observed at both integer ($\nu=-2,-1,1,2$) and fractional ($\nu=\frac{1}{3}$, $\frac{2}{3}$) moir\'e filling; for example, the incompressible peak corresponding to the $C=-4$ state at $\nu=-1$ is shown inset to Fig.~\ref{fig:3}a. To determine the Chern number of these states, we measure the $B_z$-dependence of the density, $n_e$ at which the incompressible peak appears. 
For a gapped state, the density changes according to the Streda formula\cite{streda_quantised_1982}.  

Fig.~\ref{fig:3}b shows this measurement at $\nu=-1$; 
as expected from the measured quantized Hall conductance,
the incompressible peak disperses linearly with a slope consistent with $C=-4$.
Applying this to other correlated insulators, we find the $\nu=1$ state for $D>0$  (i.e., the layer polarization corresponding to a weak moir\'e potential) is a Chern insulator with $C=1$ persisting to $B=0$. 
The adjacent $\nu=2/3$ state shows $C=2/3$, making it a fractional Chern insulator~\cite{park_observation_2023,lu_fractional_2024, xie_even-_2024}.  
As summarized in Fig.~\ref{fig:3}d, most of the other observed incompressible phases have zero Chern number, including those at $\nu=\frac{1}{3}$ and $\frac{2}{3}$ for $D<0$ (corresponding to the regime of strong moir\'e potential). 

Our compressibility measurements allow us to quantitatively determine the thermodynamic gaps of the Chern insulator state via  the relation $\Delta \mu=\int (\kappa)\si{d}n_e$; the gap is equivalent to the area of the compressibility peaks shown in Fig.~\ref{fig:3}e.  
In a clean system, this gap corresponds to the energy difference between adding and subtracting a single electron from the incompressible ground state.  
Fig.~\ref{fig:3}f shows the temperature dependence of the thermodynamic gap $\Delta$ for the topologically nontrivial states at $\nu=\pm1$ and $\frac{2}{3}$.  
The fractional Chern insulator shows a gap of $\Delta\approx 2.7 \,\si{K}$ at the lowest temperatures.

\section{Spin-orbit induced superconductivity}

\begin{figure}
\centering
\includegraphics[width=\columnwidth]{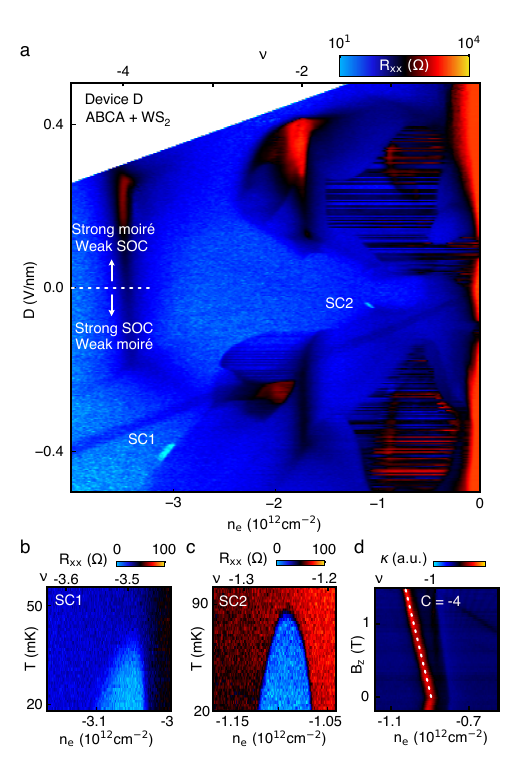}
\caption{
\textbf{Superconductivity and Chern insulator in spin-orbit coupled rhombohedral tetralayer graphene with moir\'e superlattice.} \textbf{a}, $R_{xx}$ as a function of $D$ and $n_e$. Two regions of superconductivity are observed: SC1 at higher hole density and $\vert D \vert$, and SC2 at lower $\vert n_e \vert$ and $\vert D \vert$. \textbf{b}, \textbf{c}, Superconducting domes, $R_{xx}$ as a function of $T$ and $n_e$ for SC1 at $D=-0.399 \,\si{V/nm}$ and SC2 at $D=-0.043 \,\si{V/nm}$. 
    \textbf{d}, $\kappa$ around $\nu = -1$ as a function of $B_z$ and $n_e$ taken at $D=-0.37 \,\si{V/nm}$, showing non-trivial Chern number $C=-4$.}
    \label{fig:4}
\end{figure}

Prior work on Bernal bilayer~\cite{zhang_enhanced_2023,holleis_nematicity_2024,li_tunable_2024} and rhombohedral trilayer~\cite{patterson_superconductivity_2024,yang_diverse_2024} 
graphene have shown that the domain and critical temperature of superconductivity are significantly enhanced by incorporating a transition metal dichalcogenide cladding layer.  
Notably, in these devices, spin-orbit enhanced superconductivity is observed when carriers are polarized to the graphene layer \textit{closest} to the dichalcogenide substrate~\cite{khoo_-demand_2017,gmitra_proximity_2017,island_spinorbit-driven_2019}. 
In hBN aligned rhombohedral multilayers~\cite{chen_tunable_2020,lu_fractional_2024}, Chern insulators are observed when carriers are polarized \textit{away} from the aligned hBN.  Spin-orbit enhanced superconductivity and topologically nontrivial bands are thus compatible, and may be realized by replacing the misaligned hBN substrate with a dichalcogenide layer.

Fig.~\ref{fig:4} shows $R_{xx}$ measured in Device D, which consists of a rhombohedral tetralayer sample encapsulated between WS$_2$ and hBN flakes.  The alignment of the hBN and graphene crystal lattices produces a moir\'e lattice with $\lambda\approx12 \,\si{nm}$, suggesting a small rotational misalignment. 
Here, $D\cdot n_{e}>0$ corresponds to strong spin-orbit proximity effect and weak moir\'e potential.  
We observe two zero-resistance pockets in this regime for hole doping, marked SC1 and SC2. SC1 appears in a similar parameter regime to the superconductivity in the bare, hBN encapsulated tetralayer shown in Fig. ~\ref{fig:1} 
(see Fig.~\ref{fig:SC_properties_DevD}). 
The transition temperature of SC1 is somewhat reduced in the presence of WS$_2$, showing a maximum $T_c \approx 35 \,\si{mK}$ (Fig.~\ref{fig:4}b), a $30\%$ reduction as compared to Device A. On the contrary, SC2 displays $T_c \approx 85 \,\si{mK}$ (Fig.4c), which is $50\%$ higher than the $T_c$ of the weak-moir\'e SC in Device A. SC2 has no analog in the bare rhombohedral tetralayer aligned to hBN of Fig.~\ref{fig:1}a, but bears a striking resemblance in terms of $n_e$- and $D$- range to the superconducting pocket recently discovered in WSe$_2$ supported rhombohedral trilayer~\cite{patterson_superconductivity_2024,yang_diverse_2024}, appearing near the low-$|D|$ boundary of an ordered state. 
Measurements of the in-plane critical field, moreover, show significant violation of the Pauli limit for SC1, but suppression of the in-plane critical field as compared to the Pauli limit for SC2 (see Fig.~\ref{fig:SC_properties_DevD}).  
This is consistent with the increased role of orbital depairing at low $D$ thought to limit in-plane critical fields in spin-orbit enhanced graphene superconductors~\cite{holleis_nematicity_2024}.  

The phase diagram of Device D is qualitatively similar to that of the hBN encapsulated samples.  
Although the anomalous Hall signal is not quantized at $\nu=-1$ (See Fig.~\ref{fig:AHE_DevD}, a fact we attribute to sample inhomogeneity), an incompressible state with Streda slope consistent with $C=-4$ (Fig.~\ref{fig:4}d) is observed just as in Devices A, B and C.  
Notably, thermodynamic measurements in Device D do reveal some discrepancies with the other devices. For example, Device D shows an incompressible phase at $\nu=-1$ on the strong moir\'e side that is absent in Device A, B, and C. This difference may arise either from different hBN-graphene twist angle or the proximity-induced spin-orbit coupling, although we note that the anomaly appears for the weak spin-orbit coupling sign of layer polarization. 

\section{Discussion}

Our measurements show that rhombohedral tetralayer graphene hosts all of the ingredients required for creating low disorder interfaces between superconducting states and a variety of chiral edge modes. 
We note that these features are unlikely to be unique to tetralayer graphene, and are expected to eventually be observed across a range of layer numbers. 
Our work shows that integer and fractional Chern insulators at $\nu=1$ and $\nu=2/3$---previously reported only in 5 and 6 layer samples---occur in 4 layer rhombohedral graphene as well.  At the same time, we find superconductivity---previously observed only in 2 and 3 layer systems without hBN alignment---persists to 4 layers and moreover survives a finite moir\'e potential.
This suggests a large parameter space of substrate choice and graphene-substrate alignment angle~\cite{zhang_twist-programmable_2024} within which to optimize superconducting parameters.  Even without further optimization, however, the exceptional gate tunability of rhombohedral tetralayer graphene already allows new experiments to probe the interface of superconducting and topological phases. 

\textbf{Acknowledgements.}
The authors would like to acknowledge discussions with Erez Berg and Mike Zaletel.
The work was primarily supported by the National Science Foundation under award DMR-2226850, with additional support provided by the Gordon and Betty Moore Foundation under award GBMF9471. 
CLP acknowledges support by the Department of Defense (DoD) through the National Defense Science and Engineering Graduate (NDSEG) Fellowship Program.
OIS acknowledges direct support by the National Science Foundation through Enabling Quantum Leap: Convergent Accelerated Discovery Foundries for Quantum Materials Science, Engineering and Information (Q-AMASE-i) award number DMR-1906325; the work also made use of shared equipment sponsored by under this award. 
KW and TT acknowledge support from the JSPS KAKENHI (Grant Numbers 21H05233 and 23H02052) and World Premier International Research Center Initiative (WPI), MEXT, Japan.

\textbf{Author Contributions.}
YJC fabricated devices A-C with the help of YSC and HS. 
CP fabricated device D with the help of XC. 
YJC, YSC, and MV performed transport and capacitance experiments on devices A-C, assisted by LFWH. MV, YSC and LFWH measured device D, assisted by CLP and OIS. 
TT and KW provided the hexagonal boron nitride crystals.  
YJC, YSC, MV, and AFY analyzed the data and wrote the paper.  
All coauthors reviewed the manuscript prior to submission.

\bibliographystyle{naturemag}
\bibliography{references}



\newpage
\onecolumngrid

\setcounter{equation}{0}
\setcounter{figure}{0}
\setcounter{table}{0}
\setcounter{page}{1}
\setcounter{section}{0}
\makeatletter
\renewcommand{\theequation}{S\arabic{equation}}
\renewcommand{\thefigure}{ED\arabic{figure}}
\renewcommand{\thepage}{\arabic{page}}


\section{Materials and Methods}

\subsection{Sample fabrication}
Rhombohedral graphene, hBN and WS$_2$ flakes, are prepared by mechanical exfoliation on SiO$_2$/Si substrates. 
The identification of rhombohedral domain is accomplished through Raman spectroscopy using 488 nm wavelength, then verified with Bruker photothermal AFM-IR spectroscopy with finer resolution. 
The rhombohedral domain is subsequently isolated by AFM anodic oxidation to prevent conversion of rhombohedral to Bernal stacking order~\cite{li_electrode-free_2018, zhou_half-_2021}. 

The stacking process is divided into two steps. 
First, for devices A-C, hBN is picked up with a poly(bisphenol A carbonate) (PC) film and used to sequentially pick up a graphite bottom gate, hBN, and graphite (for contact material). 
The stack is then flipped with gold-coated PDMS-assisted flipping technique~\cite{kim_imaging_2023}, resulting in a clean bottom part. 
Second, a graphite exfoliated on Polydimethylsiloxane (PDMS) is dropped on hBN, then picked up with suspended PC at $120 \,\si{~^{\circ} C}$, followed by the pick up of the rhombohedral graphene. 
Here we intentionally tried to align the straight edges of the hBN and the rhombohedral graphene in order to create the moir\'e superlattice.
This top part is then dropped onto the bottom part and the PC is melted at $180 \,\si{~^{\circ}C}$.  PC is washed with N-Methylpyrrolidone (NMP).
After stacking, standard e-beam lithography and reactive-ion etching are utilized and Cr/Au is deposited for electrodes. 

For device D, WS$_2$ layer is added for the bottom part to be at the top surface. 
In addition, when a rhombohedral graphene is picked up from the substrate and put down to the bottom stack, a negative 3-5 V was applied to the top gate while the silicon substrate and the rhombohedral graphene are grounded. 
The electrical connection to the top gate and the rhombohedral graphene is achieved by making an electrode on PC with gold stamping technique~\cite{kim_imaging_2023}, before picking up the rhombohedral graphene. 
This strategy aims to promote hole-doping of the graphene, a condition known to favor rhombohedral stacking order~\cite{li_global_2020}.

\subsection{Device characterization and measurements}
\begin{figure*}
    \centering
    \includegraphics{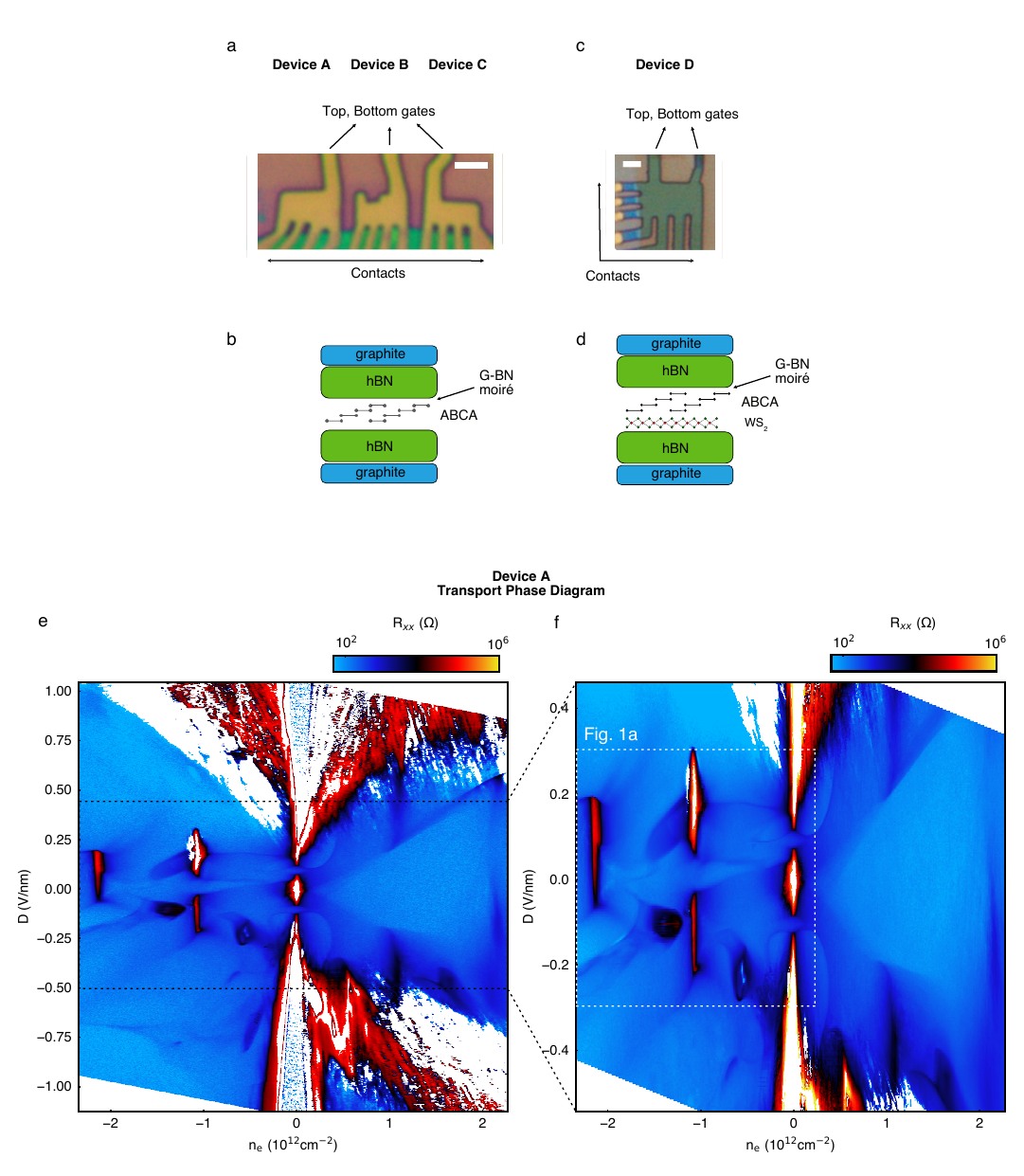}
        \caption{\textbf{Optical images and schematics of the devices, and large range transport phase diagram from Device A.} {\textbf{a}}, Optical microscope image of the ABCA devices A, B, and C. They all share the common top and bottom gates, hence the penetration capacitance measurement sums the signals from the three devices. Device A is where the main text transport data are taken. Scale bar: $2 \,\si{\mu m}$. {\textbf{b}}, Schematic of the devices A-C. {\textbf{c}}, Image of the ABCA/WS$_2$ device D with the scale bar of $2 \,\si{\mu m}$. {\textbf{d}}, Schematic for device D. {\textbf{e, f}}, Large range transport phase diagrams from device A. Electrical contact issues prevent measuring states at high $|D|$. The range where Fig.~\ref{fig:1}a is taken is outlined in \textbf{f}. At charge neutrality, we observe an insulating phase at high $|D|$ associated with a layer-polarized state and a distinct insulator near $D=0$ associated with a layer-antiferromagnet~\cite{weitz_broken-symmetry_2010,bao_stacking-dependent_2011, velasco_transport_2012, liu_spontaneous_2023, han_correlated_2023}. Additional insulating states at $\nu=-2$ are also observed, arising from the spontaneous formation of isospin polarized correlated insulating states~\cite{chen_evidence_2019}.
        }
    \label{fig:optical}
\end{figure*}

Electrical transport and penetration field capacitance measurements are performed in a dilution refrigerator, at the base temperature of $\sim 15 \,\si{mK}$ unless otherwise specified. Notably, compressibility measurements are sensitive to impedances on the $Z\sim 100\,\si{M\Omega}$ scale, making them tolerant to the high contact resistances in our devices at large $D$, which preclude transport measurements in this regime (see Fig.~\ref{fig:optical}e-f).
Four-terminal resistance is measured using lock-in amplifiers (Stanford Research Systems, SR860) with current amplifier (Basel Precision Instruments, SP983), with typical AC excitations of $0.5 - 2 \,\si{nA}$ at $17.777 \,\si{Hz}$. 
$n_{e} = (c_{tg} V_{tg} + c_{bg} V_{bg})/e$ and $D = (c_{tg} V_{tg} - c_{bg} V_{bg})/2\epsilon_{0}$ are calibrated by finite magnetic field Landau levels at electron doped side ($n_{e} > 0$) around $D = 0 \,\si{V/nm}$, where $c_{tg}$ and $c_{bg}$ are the geometric capacitances per area of the top and bottom gates, $e$ is the elementary charge, and $\epsilon_{0}$ is the vacuum permittivity. 

To measure the penetration field capacitance of the sample, we utilized high electron mobility transistor (HEMT) as described in Ref.~\cite{holleis_nematicity_2024}. Calibration of the capacitance into inverse compressibility $\kappa = \partial\mu/\partial n_{e}$ is achieved by balancing the capacitance circuit when the sample is at two different limits (metallic and insulating) as references, as described in Ref.~\cite{zibrov_tunable_2017}.
Uncalibrated $\kappa$ is sometimes used when the interest of the measurement is only to reveal the contrast between different states. 
We limit the bias current through the HEMT in order to minimize the Joule heating, especially when it matters for measuring temperature dependencies of CI and FCI (Fig.~\ref{fig:3}e,f). 
We checked that the superconducting pocket is not affected with HEMT operating, implying the effective temperature is still close to the base temperature. 
Note that since devices A-C share the same top and bottom gates (see Fig.~\ref{fig:optical}), our measurement of capacitance sums all three devices. 
However, the phase diagrams of individual devices from transport are nearly identical (see~Fig.\ref{fig:ABCA_deviceB_DeviceC}), hence mainly improving the signal to noise ratio in capacitance. 

Due to the device geometry, Hall resistance measured by the configurations $R_{13,24}$ and $R_{24,13}$ shown in Fig.~\ref{fig:Onsager} often capture $R_{xx}$ components. 
However, the measured $R_{13,24}$ is equal to $R_{xy}$ in the quantum anomalous Hall regime, showing good quantization without any symmetrization process since $R_{xx}$ drops to zero in the regime.
We plot the Onsager (anti-)symmetrized $R_{xx} = (R_{14,23} + R_{23,14})/2$ and $R_{xy} = (R_{13,24} - R_{24,13})/2$ only when finite $R_{xx}$ and geometric mixing hinders clear explanation (Fig.~\ref{fig:1}d-f, Fig.~\ref{fig:CI_energetics}a,b,e, and Fig.~\ref{fig:QAH_nD_stability}). 
We plot the raw resistances from configurations $R_{xx} = R_{14,23}$ and $R_{xy} = R_{13,24}$ in the rest of the paper. When taking magnetic field hysteresis for QAH at $\nu=-1$, 
$n_\mathrm{e}$ is tuned with $B_\mathrm{z}$ to compensate the slope from the non-zero Chern number and to stay in the QAH state (Fig.~\ref{fig:1}e,f, Fig.~\ref{fig:CI_energetics}a,b, and Fig.~\ref{fig:Onsager}.

Entering a valley-polarized state from a state with no polarization causes line-by-line switching events in two-dimensional data, as can be seen in  Fig.4a, Fig.ED3a-c, Fig.ED4b, Fig.ED5a, and Fig.ED8a. This behavior can be attributed to magnetic domains, as occur when ferromagnetic states are `zero-field cooled'. 
 Since states of opposite valley polarization are degenerate at zero magnetic field, entering a valley polarized state as a function of gates is expected to lead to an arbitrary choice of one of the degenerate configurations. 
 As in ferromagnets, other, multi-domain configurations may also be close enough in energy that they can be stabilized if the transition is crossed faster than the relevant relaxation time. Notably, states with different valley polarization have different $R_{xy}$, while multi-domain states will have different $R_{xx}$ as well.  As a result, repeated quenching into the polarized state (as happens between lines) can stabilize states with different measured transport properties, which resemble line-by-line switching noise in the data. This switching noise reflects different microscopic polarizations of the valley moments, and is suppressed by applied magnetic field, which favors a uniform polarization into a single valley.

\subsection{Pauli-limit violation ratio of ABCA and ABCA/WS$_2$ devices}
We compare the Pauli-limit violation ratio (PVR) between the hBN encapsulated ABCA (device A) and spin-orbit proximitized ABCA/WS$_2$ device (device D). The measured $B_{c,\parallel}$ at the base temperature is divided by the Pauli-limit $B_{p} = 1.86 \times T_{c}$ ($T_c$ taken at zero field) for BCS superconductor, and define PVR $= B_{c,\parallel}/B_{p}$. The raw data for the extraction can be found in Fig.~\ref{fig:SC_properties}b and c for SC in ABCA, and Fig.~\ref{fig:4}b,c, Fig.~\ref{fig:SC_properties_DevD}d,f for SC1 and SC2 in ABCA/WS$_2$. 

Without spin-orbit coupling, PVR from SC in ABCA (Fig.~\ref{fig:SC_properties}g) is around 1 or less as a function of doping, showing agreement with BCS theory. SC1 from ABCA/WS$_2$ (which occurs at a similar position to SC in ABCA) shows significantly improved PVR due to the spin-orbit coupling (Fig.~\ref{fig:SC_properties_DevD}e). Interestingly, SC2 from ABCA/WS$_2$ does not show an improvement of PVR (Fig.~\ref{fig:SC_properties_DevD}k).

\section{Extended Data}

\begin{figure*}
    \centering
    \includegraphics{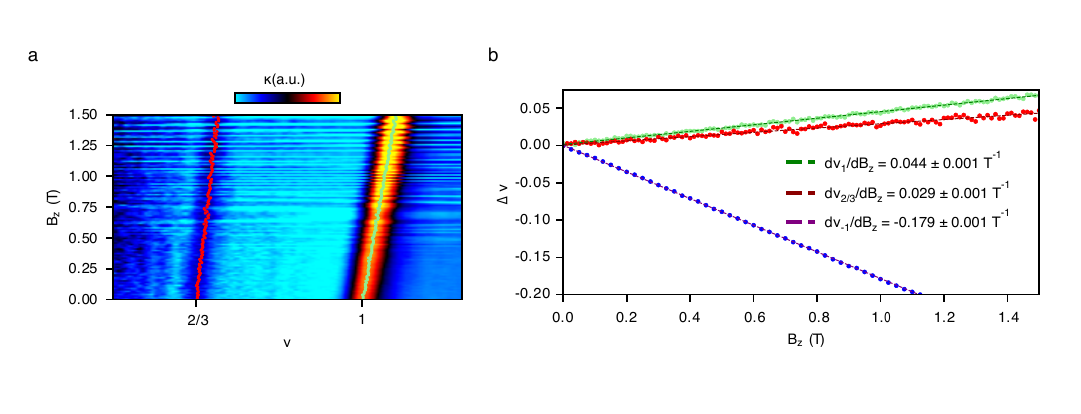}
    \caption{\textbf{Extraction of Chern numbers from capacitance.} 
    \textbf{a}, Magnetic field dependence of $\kappa$ for $\nu=1$ and $2/3$ at $D = 0.885 \,\si{V/nm}$. The red dots denote the peak position corresponding to the incompressible state appearing at $\nu = 2/3$ when B=0, while the green dots correspond to the peak position for the $\nu=1$. 
    \textbf{b}, Variation of the moire filling, $\Delta \nu(B)$ for the three incompressible states as determined from the data plotted in panel \textbf{a} and Fig.~3\textbf{b} . The dashed lines represent linear fits, with the slopes indicated in the legend.   
    The Chern numbers can be extracted from the Streda formula $C=\frac{\Phi_0}{A_{uc}}\frac{d\nu}{dB} = \Phi_{0} \frac{|n_{\pm4}|}{4}\frac{d\nu}{dB}$; with $A_{uc}$ the area of moire unit cell and $n_{\pm4}$ the carrier density at $\nu = \pm 4$. The obtained $C (\nu = 1)=0.98\pm0.03$, $C(\nu = 2/3)=0.64\pm0.03$, and $C(\nu = -1)=-4.0\pm0.1$ are consistent with the expected Chern numbers 1, 2/3, and -4, respectively, assuming $|n_{\pm4}|=(2.15 \pm 0.05) \times10^{12} \,\si{cm^{-2}}$ (determined by quantum oscillations at $D=0$). Fractional Chern insulator were also found in~\cite{spanton_observation_2018,xie_fractional_2021,cai_signatures_2023,xie_even-_2024} }
    \label{fig:slope_extraction}
\end{figure*}

\begin{figure*}
    \centering
    \includegraphics{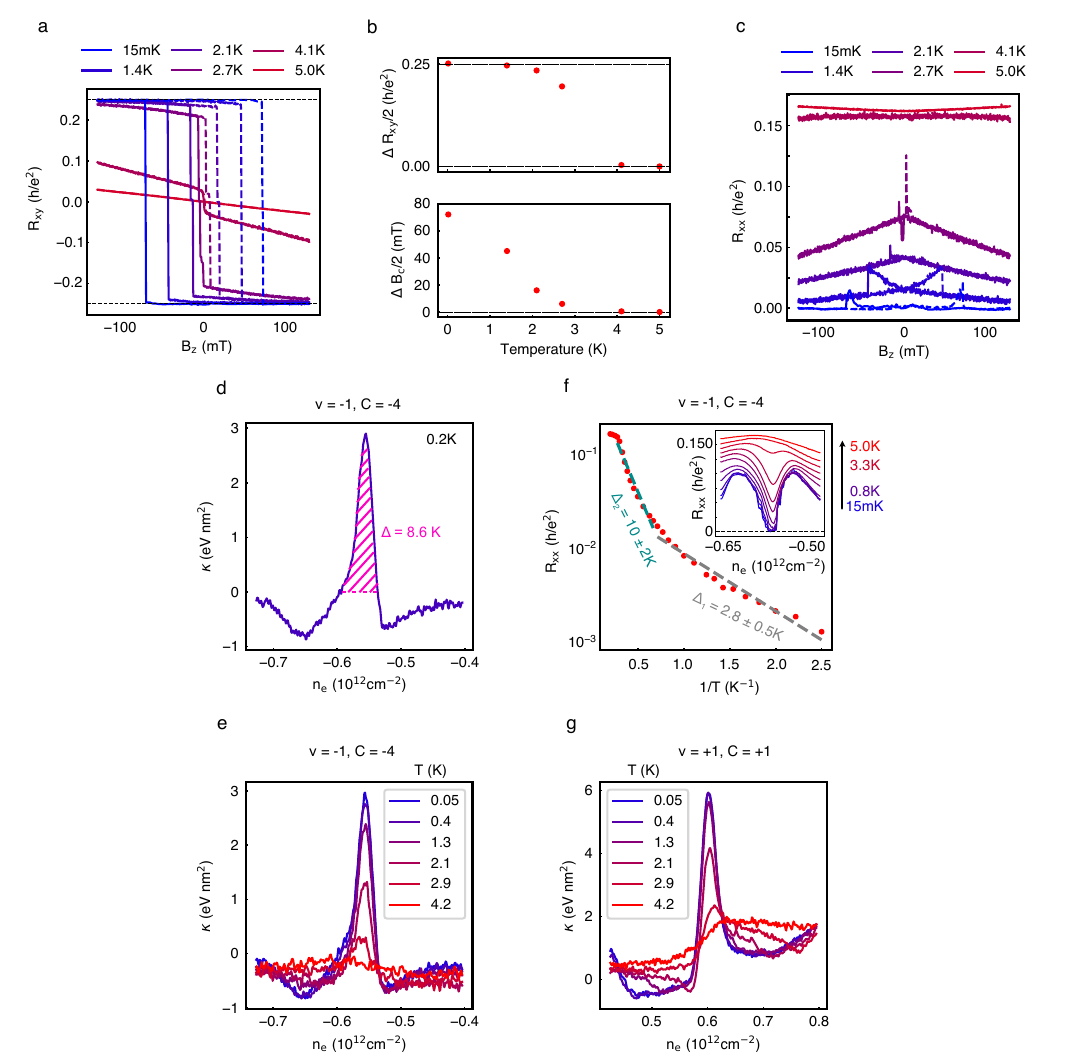}
    \caption{\textbf{Energetics of QAH at $\bm{\nu=-1}$ and $\bm{\nu=+1}$.} \textbf{a-c},  Temperature dependence of hysteresis loops of $R_\mathrm{xy}$ (\textbf{a}) and $R_\mathrm{xx}$ (\textbf{c}) taken at $n_\mathrm{e} = -0.564 \times10^{12} \,\si{cm^{-2}}$ (when $B_\mathrm{z} = 0$), showing the Curie temperature $T_\mathrm{Curie} \approx 4.5 \,\si{K}$. Panel \textbf{b} summarizes $\Delta R_\mathrm{xy}/2$ ($B_\mathrm{z} = 0$) (upper) and coercive field (lower) from \textbf{a} as a function of temperature for clarity. \textbf{d}, Gap determination from $\kappa$, showing the case of $\nu=-1$ Chern insulator. We integrate $\kappa = \partial \mu/\partial n_e$ above zero over the incompressible peak to estimate the chemical potential jump. \textbf{e}, $T$ dependence of $\kappa$ for $\nu=-1$ Chern insulator. \textbf{f}, Activation gap measurement of the QAH state at $\nu = -1$. The inset shows the temperature dependence of $R_{xx}$ as a function of $n_e$ around the quantized region of QAH. The dip at low temperature corresponding to QAH regime fills up as $T$ increases. The main panel shows temperature dependence of $R_\mathrm{xx}$ inside of the dip (at the same position as \textbf{a} and \textbf{c}). The activation gap $\Delta$ can be obtained by the Arrhenius fitting following $R_{xx} \sim e^{-\Delta/(2T)}$. We find around a decade of linear activation for different temperature ranges, which gives $\Delta_{1} = 2.8 \pm 0.5 \,\si{K}$ for low temperature and $\Delta_{2} = 10 \pm 2 \,\si{K}$ for intermediate temperature regimes. $\Delta_{2}$ agrees better to the measured thermodynamic gap from \textbf{d}. We interpret the smaller gap ($\Delta_1$) extracted at lower temperature as arising from disorder-mediate hopping\cite{shklovskii_electronic_1984}.} \textbf{a}-\textbf{f} are taken at $D = -0.208 \,\si{V/nm}$.
    \textbf{g}, $T$ dependence of $\kappa$ for $\nu=+1$ Chern insulator at $D = 0.835 \,\si{V/nm}$. 
    \label{fig:CI_energetics}
\end{figure*}

\begin{figure*}
    \centering
    \includegraphics{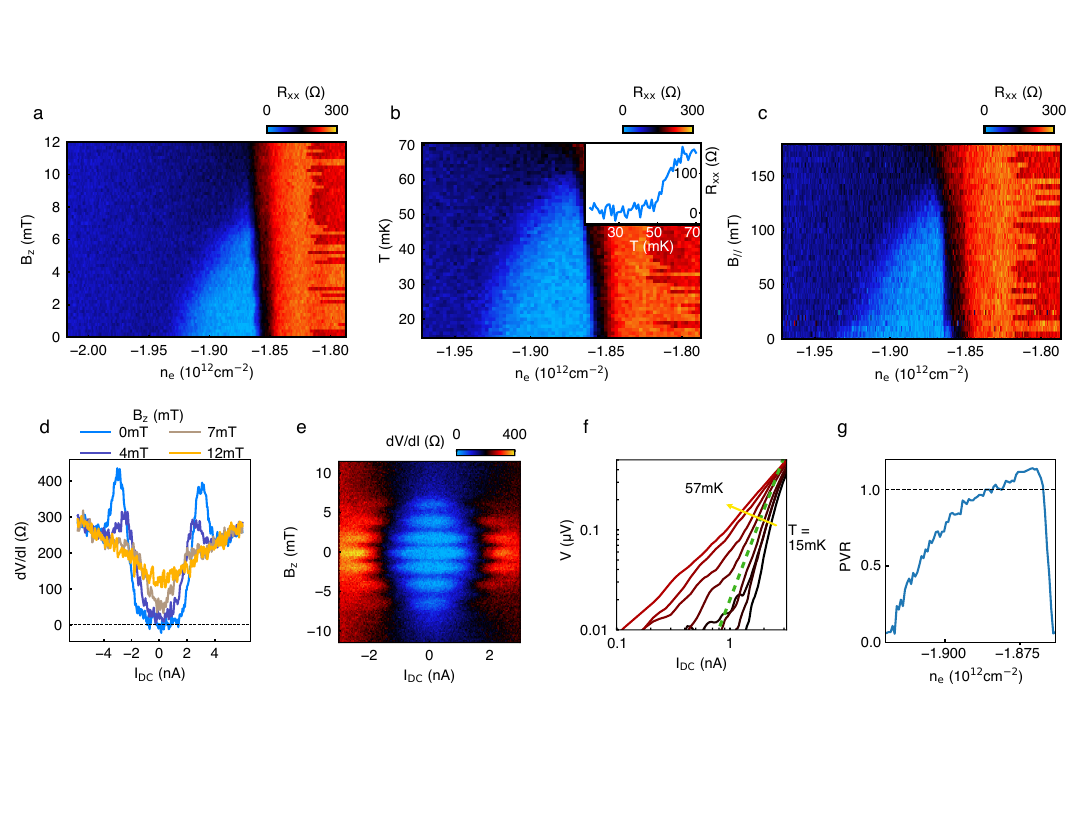}
    \caption{\textbf{Characterization of the superconductivity.} \textbf{a-c}, $R_{xx}$ as a function of electron density $n_e$ and out-of-plane field $B_z$ (\textbf{a}), temperature $T$ (\textbf{b}), and in-plane field $B_{\parallel}$ (\textbf{c}). Inset of \textbf{b} shows $R_{xx}$ vs T at the optimal $n_e = -1.868 \times10^{12} \,\si{cm^{-2}}$. \textbf{d}, $B_z$ dependent $\si{d}V/\si{d}I$ vs $I_{DC}$ at $n_e = -1.879 \times10^{12} \,\si{cm^{-2}}$, $D = -0.138 \,\si{V/nm}$. \textbf{e}, $B_z$ dependent $\si{d}V/\si{d}I$ at $n_e = -1.865 \times10^{12} \,\si{cm^{-2}}$, close to the right-side boundary of the superconducting pocket. Oscillation due to macroscopic interference is observed, corroborating coherence of the superconducting state. \textbf{f}, Temperature dependent $V$ vs $I_{DC}$ taken at $n_e = -1.879 \times10^{12} \,\si{cm^{-2}}$. The green dashed line indicates where $V \propto I^3$, showing $T_{BKT} \approx 40 \,\si{mK}$. \textbf{g}, Pauli limit violation ratio (PVR) as a function of $n_e$, showing overall obedience of the Pauli limit (see Methods for the discussion). All data here are taken at $D = -0.138 \,\si{V/nm}$.}
    \label{fig:SC_properties}
\end{figure*}

\begin{figure*}
    \centering
    \includegraphics{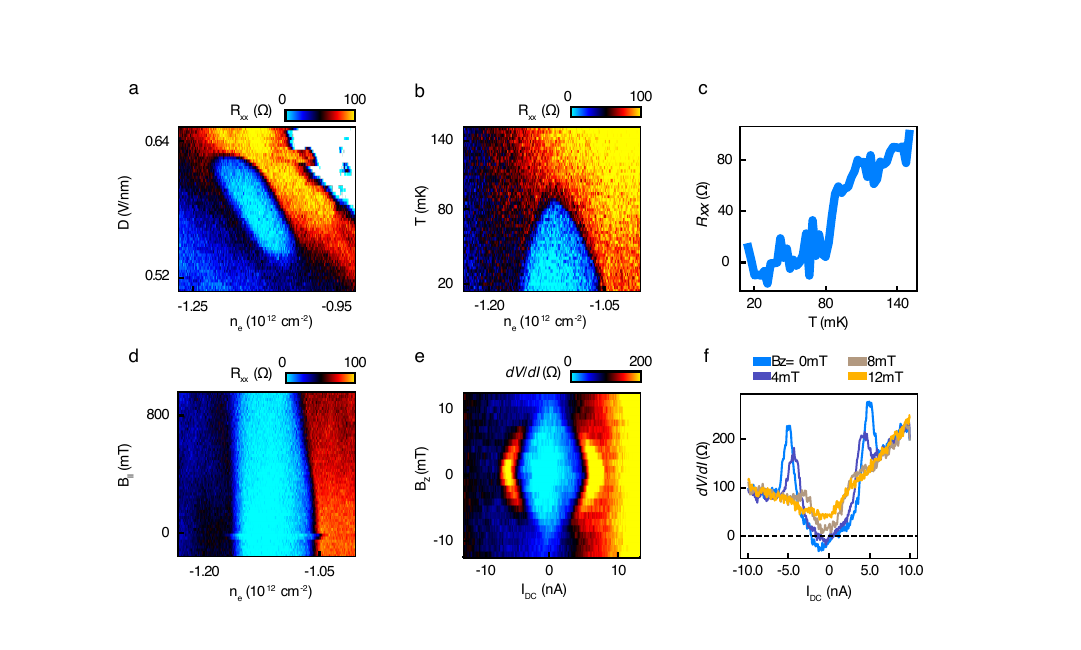}
    \caption{\textbf{Characterization of superconductivity for strong moir\'e potential Device A.} 
    \textbf{a}, ($n_e$, $D$) phase diagram of $R_{xx}$ around the superconducting pocket on the strong moiré side. The white region of the phase diagram indicates where the contact resistance becomes too large for transport measurements. \textbf{b}, $R_{xx}$ as a function of $T$ and $n_e$ taken at $D=0.576 \,\si{V/nm}$. 
    \textbf{c}, $R_{xx}$ vs $T$ at $n_e = -1.125 \times10^{12} \,\si{cm^{-2}}$. 
    \textbf{d}, In plane field dependence, $R_{xx}$ as a function of $B_{\parallel}$ and $n_e$ taken at $D=0.576 \,\si{V/nm}$. 
    \textbf{e, f}, $\si{d}V/\si{d}I$ as a function of $B_z$ and $I_{DC}$ at $n_e = -1.123 \times10^{12} \,\si{cm^{-2}}$ and $D=0.576 \,\si{V/nm}$.  The measured critical current $I_c \approx 5 \,\si{nA}$, while the critical out-of-plane field $B_c$ is $10 \,\si{mT}$.}
    \label{fig:spin_pol}
\end{figure*}

\begin{figure*}
    \centering
    \includegraphics{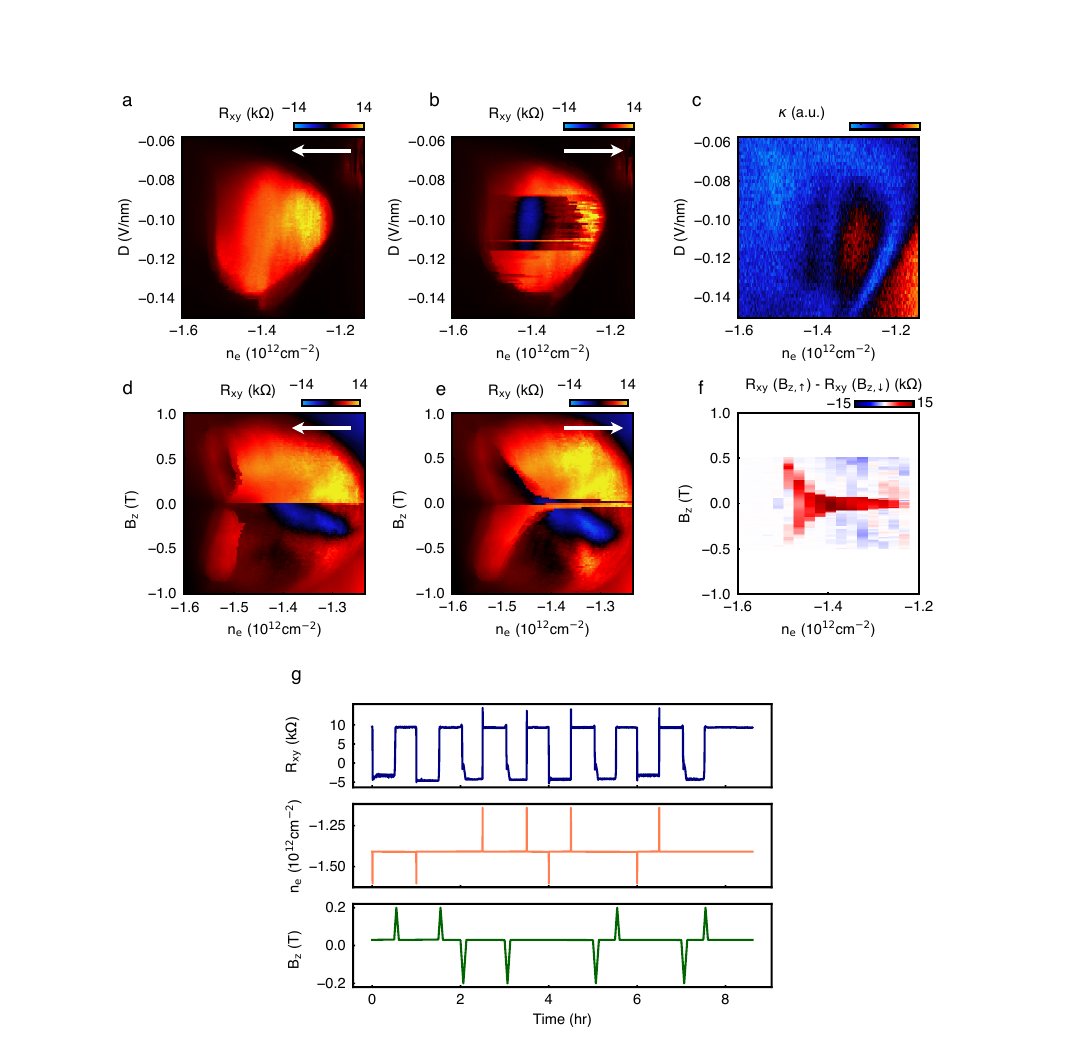}
    \caption{\textbf{Anomalous Hall effect and electrical switching around $\bm{\nu \sim -2.5}$.} \textbf{a}, ($n_e$, $D$) dependent $R_{xy}$ at $B_z = 30 \,\si{mT}$ with $n_e$ as the fast sweep axis from right to left. \textbf{b}, same as (\textbf{a}) but with the opposite sweep direction. \textbf{c}, $\kappa = \partial \mu/ \partial n_{e}$ from the penetration capacitance at the same range as \textbf{a} and \textbf{b}. \textbf{d}, \textbf{e}, ($n_e$, $B_z$) dependent $R_{xy}$ with different sweep directions in $n_e$. The magnetic moment $m$ of the state changes sign around $n_e = -1.5 \times10^{12} \,\si{cm^{-2}}$ when $B_z$ is around zero, and the mechanism of the electrical switch is similar to the mechanism discussed in Fig.~\ref{fig:2}. \textbf{f}, Resistance difference between $R_{xy}$ when sweeping up in $B_z$ and $R_{xy}$ sweeping down in $B_z$. \textbf{g}, Non-volatile switching of the two states, controlled by $B_z$ and $n_e$. \textbf{d-g} are taken with $D = -0.102 \,\si{V/nm}$. All data here are taken at $200 \,\si{mK}$.}
    \label{fig:AH_nu-2.5}
\end{figure*}

\begin{figure*}
    \centering
    \includegraphics{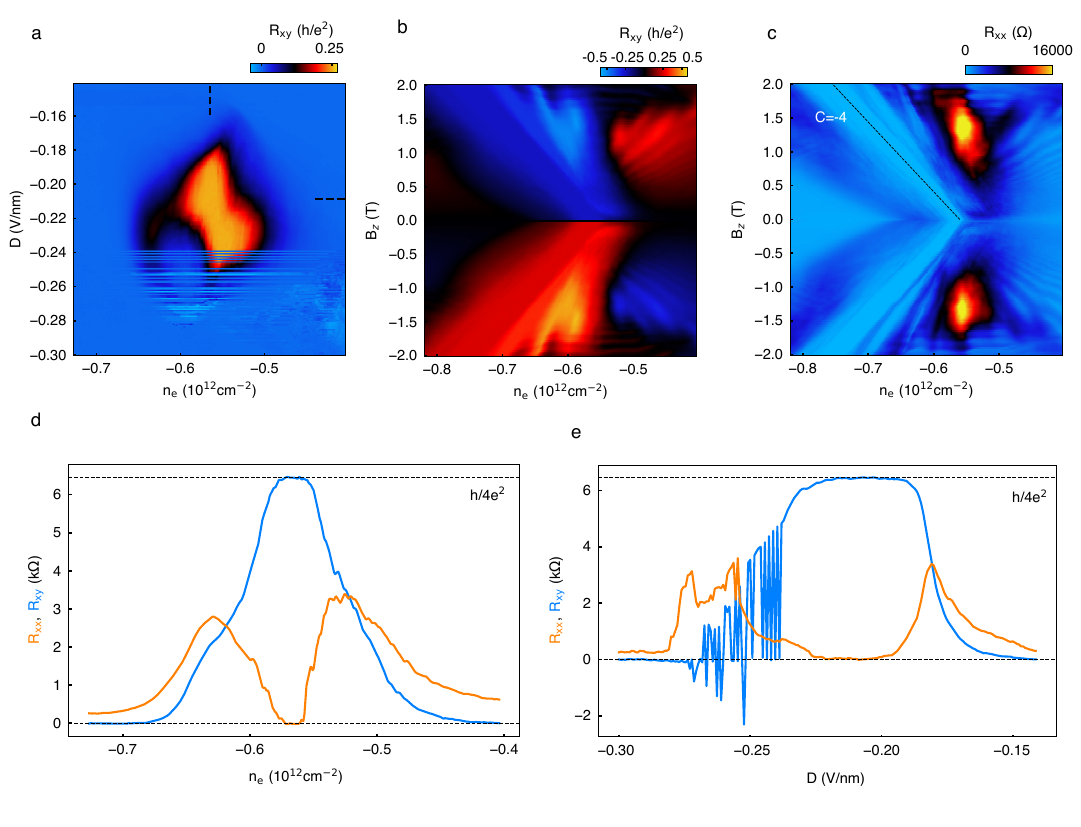}
    \caption{\textbf{Landau fans and stability of quantization in QAH state at $\nu=-1$.} \textbf{a}, ($n_e$, $D$) dependent $R_{xy}$ around QAH state with $n_e$ as the fast sweep axis. The switching behavior indicates the closeness of the two states in energy. The dashed lines are the positions where \textbf{d} and \textbf{e} are taken.  \textbf{b}, $R_{xy}$ Landau fan, showing a plateau along the $\vert C \vert = 4$ slope. \textbf{c}, $R_{xx}$ Landau fan, where the dashed line correspond to $\vert C \vert = 4$ from Streda formula. Both fans are taken at $D = -0.208 \,\si{V/nm}$. \textbf{d}, Linecuts with constant $D = -0.208 \,\si{V/nm}$, showing quantized $R_{xy}$ around the value $h/4e^2$ and $R_{xx}$ around zero. \textbf{e}, Linecuts with constant $n_e = -0.564 \times10^{12} \,\si{cm^{-2}}$.}
    \label{fig:QAH_nD_stability}
\end{figure*}

\begin{figure*}
    \centering
    \includegraphics[width = 183mm]{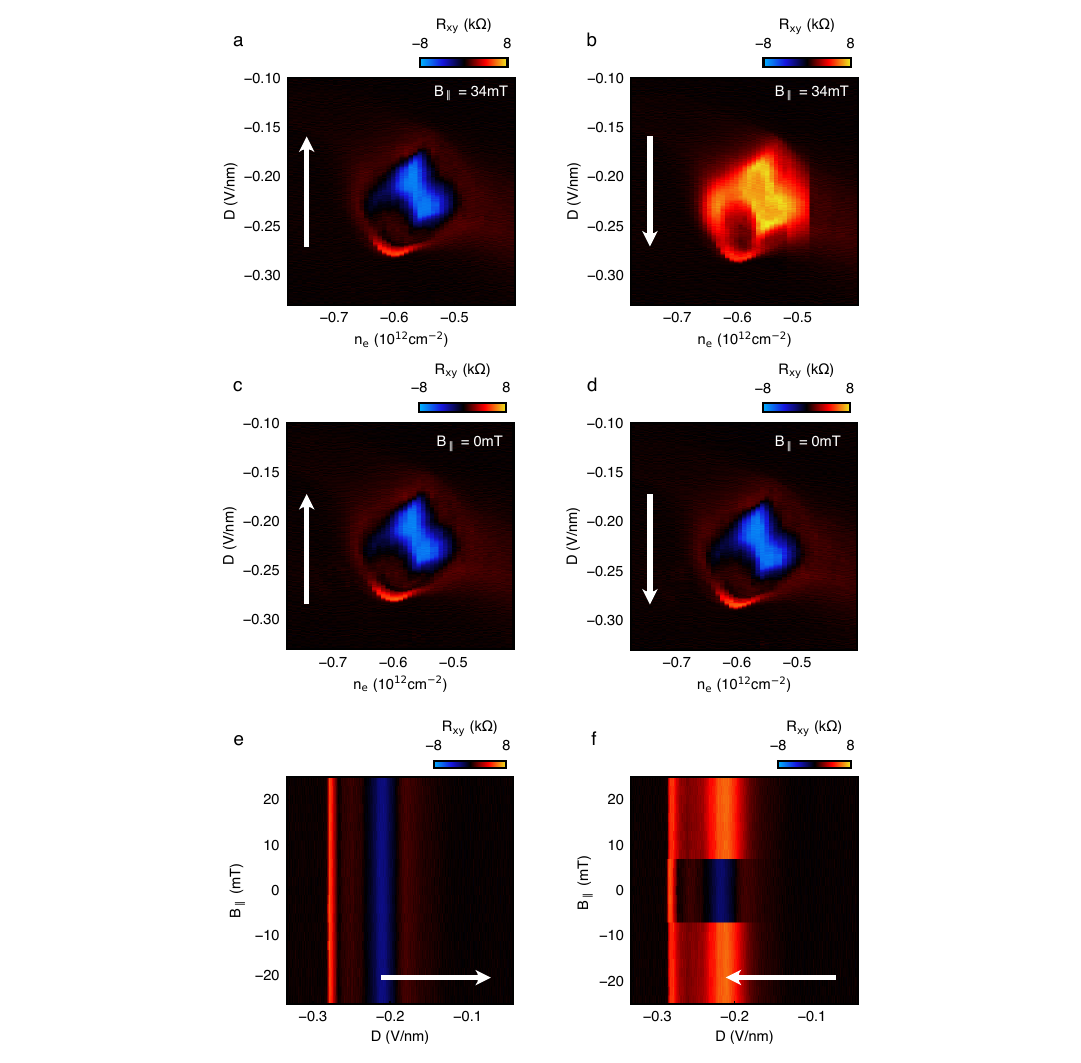}
    \caption{\textbf{In-plane magnetic field dependence of electric switching at $\nu = -1$.} 
    \textbf{a}, \textbf{b}, ($n_e$, $D$) dependent $R_{xy}$ at $B_\parallel = 34 \,\si{mT}$, $D$ as the fast sweep axis. Sweep direction is indicated in the arrows. At sufficiently large $B_\parallel$, the electrical switching exists. 
    \textbf{c}, \textbf{d}, Same as \textbf{a} and \textbf{b}, but $B_\parallel = 0 \,\si{mT}$. The switching is not present. 
    \textbf{e}, \textbf{f}, $B_\parallel$ dependence of the switching at a fixed $n_e = -0.607 \times10^{12} \,\si{cm^{-2}}$. A small but finite $B_\parallel$ is required in order to observe switching. The data here are taken with nominal $B_z = 5\,\si{mT}$.
    }
    \label{fig:inplane_switching}
\end{figure*}

\begin{figure*}
    \centering
    \includegraphics[width = 173mm]{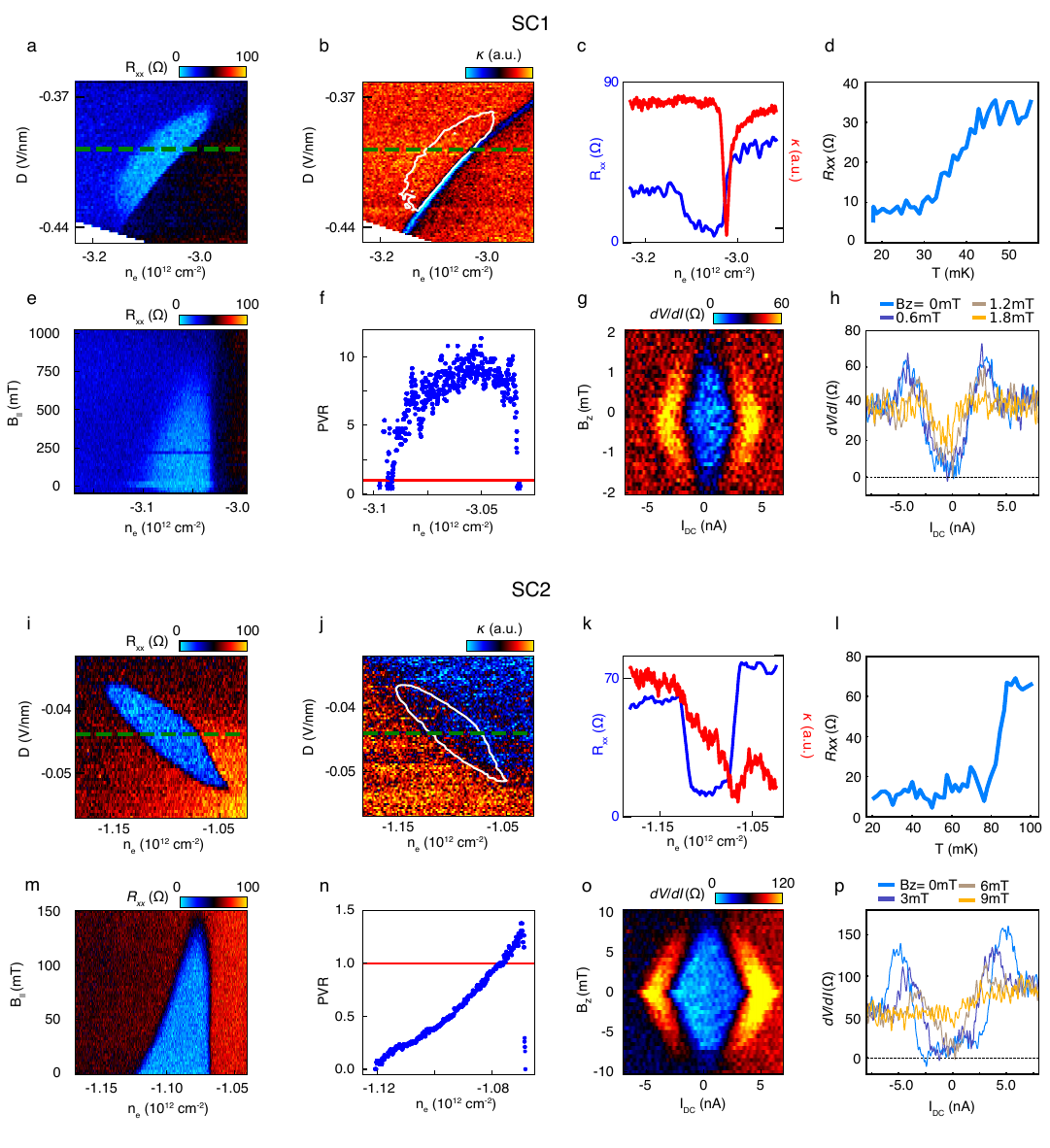}
    \caption{\textbf{Characterization of superconductivity SC1 and SC2 in ABCA/WS$_2$ device D.} \textbf{a}, ($n_e$, $D$) phase diagram of $R_{xx}$ around the superconducting pocket SC1. \textbf{b}, Inverse compressibility $\kappa$ measured around SC1. White contour indicates the superconducting region extracted from \textbf{a}. The dashed line is at $D=-0.399 \,\si{V/nm}$, where \textbf{c}-\textbf{h} are taken. \textbf{c}, Linecuts of $R_{xx}$ (blue) and $\kappa$ (red) for SC1, where a clear dip of $\kappa$ is visible on the right side of the superconducting region suggesting a first-order phase transition. \textbf{d}, $R_{xx}$ vs $T$ at $n_e = -3.049 \times10^{12} \,\si{cm^{-2}}$. \textbf{e}, $B_{\parallel}$ dependence of SC1. \textbf{f}, PVR for SC1 as a function of doping, showing a huge improvement compared to the case of ABCA device without WS$_2$. \textbf{g, h}, $\si{d}V/\si{d}I$ as a function of $B_z$ and $I_{DC}$ at $n_e = -3.051 \times10^{12} \,\si{cm^{-2}}$, manifesting critical current $I_c \approx 2 \,\si{nA}$ and critical field $B_c$ around $1.5 \,\si{mT}$. \textbf{i}, \textbf{j}, $n_e$, $D$ phase diagram of $R_{xx}$ and $\kappa$ of SC2. \textbf{k}, $R_{xx}$ and $\kappa$ linecuts, with $\kappa$ showing a slight dip on the right side of the superconducting region. \textbf{l}, $R_{xx}$ vs $T$ at $n_e = -1.092 \times10^{12} \,\si{cm^{-2}}$.
    \textbf{m}, \textbf{n}, $B_{\parallel}$ dependence and PVR of SC2 showing overall obedience of the Pauli limit. \textbf{o, p}, $\si{d}V/\si{d}I$ as a function of $B_z$ and $I_{DC}$ at $n_e = -1.104 \times10^{12} \,\si{cm^{-2}}$, $D=-0.043 \,\si{V/nm}$, with $I_c \approx 3 \,\si{nA}$ and $B_c \approx 8 \,\si{mT}$.}
    \label{fig:SC_properties_DevD}
\end{figure*}

\begin{figure*}
    \centering
    \includegraphics[width = 183mm]{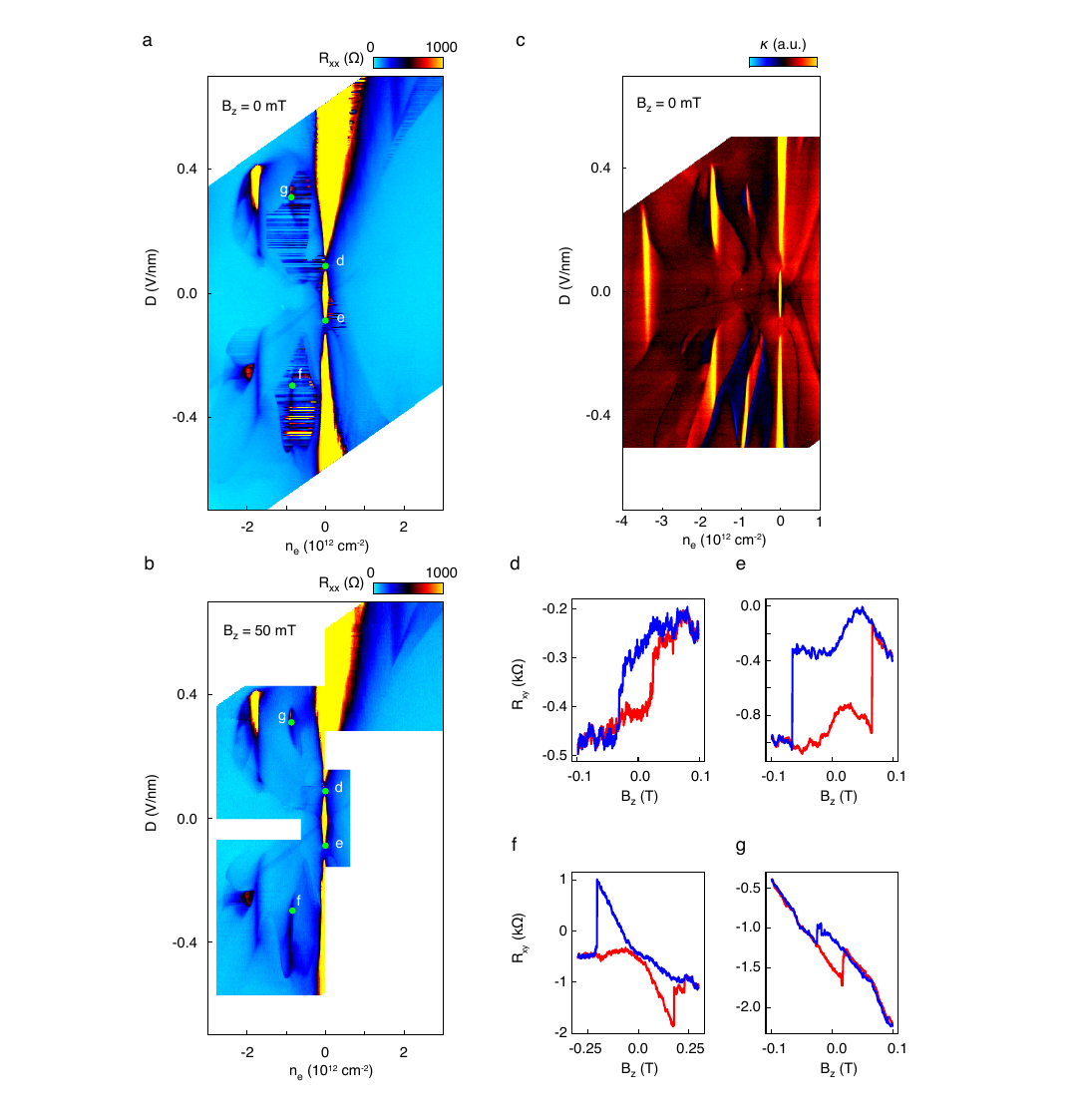}
    \caption{\textbf{Phase diagram and anomalous Hall effect in ABCA/WS$_2$ device D.} \textbf{a}, ($n_e$, $D$) dependent $R_{xx}$ at zero magnetic field. Several switchy regions are observed. \textbf{b}, Switchy behavior is suppressed by a small out-of-plane magnetic field of $B_z = 50 \,\si{mT}$.  This is consistent with a magnetic origin, with bistability caused by different orbital magnetization states. \textbf{c}, ($n_e$, $D$) dependent penetration capacitance at zero magnetic field. \textbf{d-g}, $R_{xy}$ hysteresis loops as a function of $B_z$. \textbf{d} and \textbf{e} are at $\nu = 0$ and \textbf{f} and \textbf{g} are at $\nu = -1$, as marked in \textbf{a} and \textbf{b}.}
    \label{fig:AHE_DevD}
\end{figure*}


\begin{figure*}
    \centering
    \includegraphics{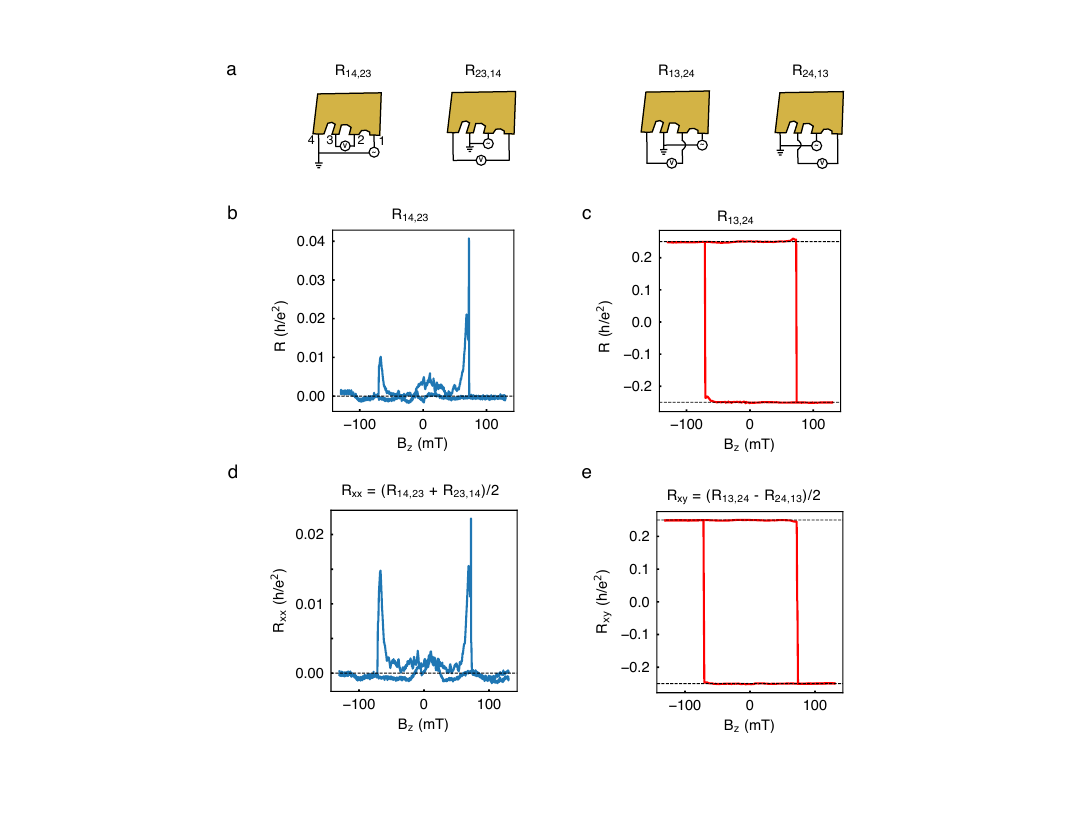}
    \caption{\textbf{Onsager (anti-)symmetrization.} \textbf{a}, Contact configuration $R_{14,23}$ and its Onsager pair $R_{23,14}$ for $R_{xx}$ measurement, and $R_{13,24}$ and its Onsager pair $R_{24,13}$ for $R_{xy}$ measurement. \textbf{b-e}, Process to obtain Fig. \ref{fig:1}e and f. Resistances from $R_{14,13}$ and $R_{13,24}$ contact configurations are shown in \textbf{b} and \textbf{c}, roughly corresponding to $R_{xx}$ and $R_{xy}$ each and already showing good quantization since $R_{xx}$ goes to zero in QAH, minimizing the geometrical mixing. \textbf{d} and \textbf{e} are obtained by Onsager symmetrization, showing less deviation from the quantized value, particularly at $B_z$ around the coercive fields where the $R_{xx}$ becomes large and the geometrical mixing presents. In the quantum anomalous Hall phase, we find $R_{xx} < 0$, which may be associated with  coupling of edge modes via localized states~\cite{buttiker_negative_1988,kaverzin_negative_2024}.}
    \label{fig:Onsager}
\end{figure*}

\begin{figure*}
    \centering
    \includegraphics{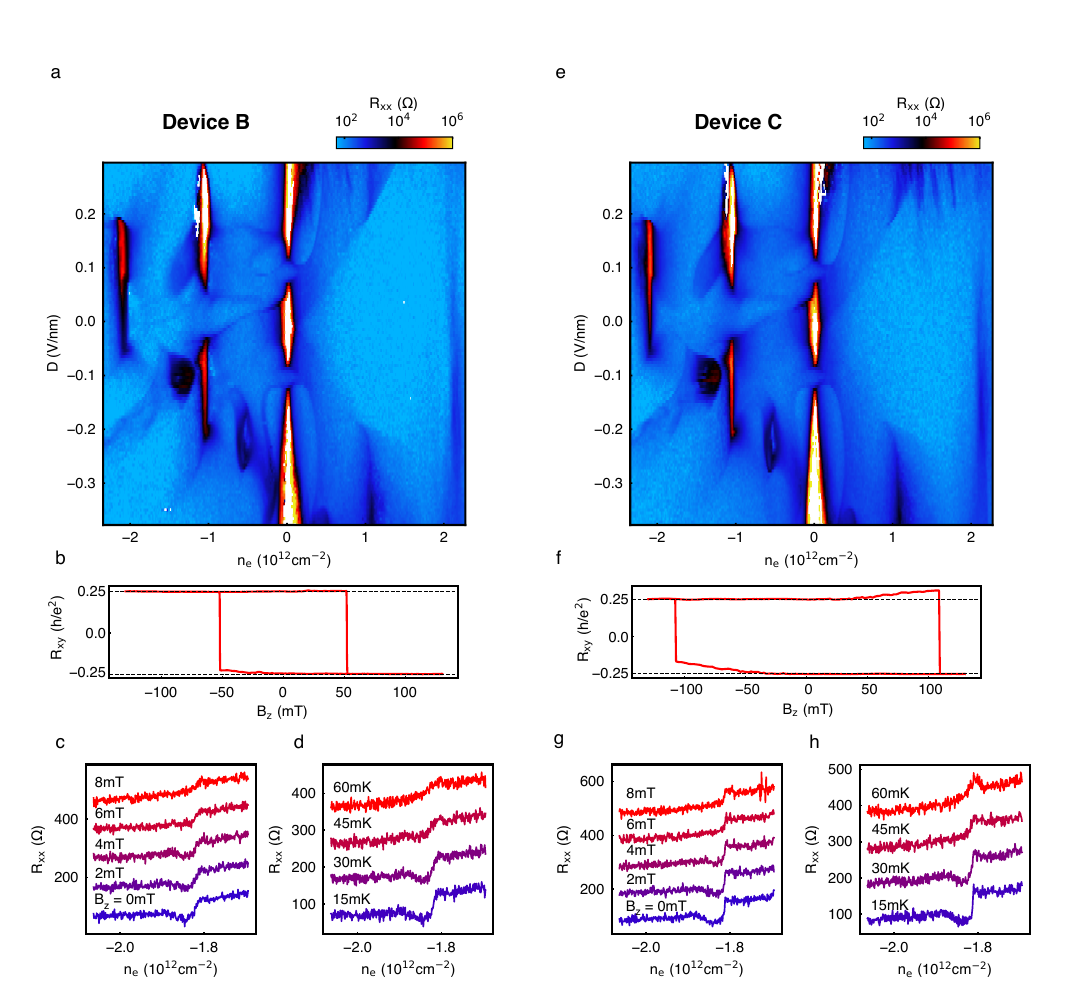}
    \caption{\textbf{Reproducibility, signatures of superconductivity and QAH states in devices B and C.} \textbf{a}, ($n_e$, $D$) dependent $R_{xx}$ in device B, showing the same features at nearly identical positions with device A. \textbf{b}, Hysteresis loop of $R_{xy}$ for the QAH state in device B taken at $n_e = -0.558 \times10^{12} \,\si{cm^{-2}}$ and $D = -0.208 \,\si{V/nm}$, showing a good quantization around $h/4e^2$. \textbf{c}, \textbf{d}, $B_z$ and temperature dependence of $R_{xx}$ along the linecut at $D=-0.131 \,\si{V/nm}$, crossing the signature of superconductivity. The critical field and temperature values from the resistance dip are similar to the superconductivity in device A, despite the fact that the resistance does not drop to zero due to the mixing of filtering grounds for different contacts in devices B and C.
    Curves are offset by 100$\Omega$ for clarity. \textbf{e-h}, Dataset from Device C. ($n_e$, $D$) dependent $R_{xx}$ (\textbf{e}), $R_{xy}$ hysteresis loop for QAH state taken at $n_e = -0.562 \times10^{12} \,\si{cm^{-2}}$ and $D = -0.208 \,\si{V/nm}$ (\textbf{f}), $B_z$ and $T$ dependence of $R_{xx}$ for the superconducting state taken at $D=-0.131 \,\si{V/nm}$ (\textbf{g}).}
    \label{fig:ABCA_deviceB_DeviceC}
\end{figure*}


\clearpage

\end{document}